\documentclass[aps, prl, twocolumn, superscriptaddress]{revtex4}
\usepackage{graphicx}
\usepackage{array}
\usepackage{amsmath,amssymb}
\usepackage{mathrsfs}
\usepackage[dvipsnames,usenames]{color}
\begin{document}

\title{Topological phase transitions of Thouless charge pumping realized in helical organic molecules with long-range hoppings}
\author{Ai-Min Guo}
\email[]{aimin.guo@csu.edu.cn}
\affiliation{Hunan Key Laboratory for Super-microstructure and Ultrafast Process, School of Physics and Electronics, Central South University, Changsha 410083, China}
\author{Pei-Jia Hu}
\affiliation{Hunan Key Laboratory for Super-microstructure and Ultrafast Process, School of Physics and Electronics, Central South University, Changsha 410083, China}
\author{Xiao-Hui Gao}
\affiliation{Hunan Key Laboratory for Super-microstructure and Ultrafast Process, School of Physics and Electronics, Central South University, Changsha 410083, China}
\author{Tie-Feng Fang}
\email[]{fangtiefeng@lzu.edu.cn}
\affiliation{Key Laboratory for Magnetism and Magnetic Materials of MOE, School of Physical Science and Technology, Lanzhou University, Lanzhou 730000, China}
\author{Qing-Feng Sun}
\affiliation{International Center for Quantum Materials, School of Physics, Peking University, Beijing 100871, China}
\affiliation{Collaborative Innovation Center of Quantum Matter, Beijing 100871, China}
\affiliation{CAS Center for Excellence in Topological Quantum Computation, University of Chinese Academy of Sciences, Beijing 100190, China}

\date{\today}

\begin{abstract}
Recent studies indicated that helical organic molecules, such as DNA and $\alpha$-helical protein, can behave as Thouless quantum pumps when a rotating electric field is applied perpendicularly to their helical axes. Here we investigate the influence of long-range hoppings on this topological pumping of electrons in single-helical organic molecules. Under variation of the long-range hoppings governed by a decay exponent $\mu$, we find an energy gap in the molecular band structure closes at a critical value $\mu_c$ of the decay exponent and reopens for $\mu$ deviating from $\mu_c$. The relevant bulk bands in a pumping cycle acquire different Chern numbers in the strong ($\mu<\mu_c$) and weak ($\mu>\mu_c$) long-range hopping regimes, with a sudden jump at criticality. This topological phase transition is also shown to separate two distinct behaviors of the midgap end states in the pumping process. The end states carry quantized current pumped by the rotating electric field and the current forms a plateau by sweeping the Fermi energy over the gap. In the strong hopping phase, the quantized current plateau is positive, which is reversed to a negative one with smaller amplitude in the weak hopping phase. However, the reversal is a smooth crossover, not a sharp transition, due to the finite sizes of the molecules. We show that these transport characteristics of the topological phase transition could also be observed at finite temperatures.
\end{abstract}

\maketitle

\section{I. Introduction}
Topology is a mathematical concept that describes some invariant properties of geometrical objects when they are smoothly deformed. The study of topological physics began with the discoveries of the integer \cite{Klitzing1980} and fractional \cite{Tsui1982} quantum Hall effects, and the gapped quantum spin-liquid state of integer-spin chains \cite{Haldane1981,Haldane1983a,Haldane1983b}. The integer quantum Hall effect defines a first topological phase that is distinct from all states of matter known before. In this effect, the quantized Hall conductance carried by chiral edge states is a topological invariant independent of material details \cite{Thouless1982,Kohmoto1985}. When the time-reversal symmetry is preserved, some spin-orbit coupled systems can support exotic topological insulating states \cite{Kane2005, Bernevig2006a, Bernevig2006b, Konig2007, Moore2009, Hasan2010, Qi2011}, which further enrich our understanding of topological phases of matter. Topological concepts can also be applied to unconventional superconductors \cite{Qi2011}, superfluids \cite{Qi2009,Chung2009,Murakawa2009}, and nodal systems \cite{Schnyder2011,Wang2012,Wan2011}. Indeed, it has now become apparent that topological phenomena are essentially a ubiquitous property of physical systems in diverse fields including condensed-matter physics, ultracold atomic gases \cite{Cooper2019}, photonics \cite{Ozawa2019}, and even classical mechanics \cite{Huber2016}. For gapped topological quantum materials, the global structure of the wave function, characterized by a topological invariant, remains unchanged under variations of system parameters, unless the gap is closed at a topological phase transition. This wave function topology manifests itself as gapless states localized at the system boundary. Interestingly, topological invariants, such as the Chern number, can count the number of particles that are pumped through a spatially periodic system that is driven periodically and adiabatically in time. This topological quantum pumping, first introduced by Thouless in 1983 \cite{Thouless1983}, has recently been demonstrated in ultracold atomic experiments \cite{Nakajima2016,Lohse2016}.

While most existing studies of topological phases of matter are based on inorganic materials, recent works \cite{Guo2017,Tang2019} have proposed that topological states can also emerge in helical organic or bioorganic molecules. A distinct feature of these molecules, such as DNA and $\alpha$-helical protein, is their unique helical structures. Due to this helical symmetry, the $\alpha$-helical protein and DNA provide direct realizations \cite{Guo2017} of the topological Thouless pump driven by a rotating electric field applied perpendicularly to their helical axes. The pumped current through these molecules exhibits quantized plateaus by sweeping the Fermi energy over band gaps, which are topologically protected against perturbations \cite{Guo2017}. It has further been shown that single-stranded DNA in the vicinity of a conventional superconductor can support topologically nontrivial superconducting phases hosting Majorana zero modes at the ends \cite{Tang2019}. These fascinating results may stimulate more research interest in exploring topological physics of bioorganic systems. On the other hand, DNA- and protein-based structures have been extensively studied as promising candidates for functional nanoelectronic devices \cite{Endres2004, Meredith2013, Albuquerque2014, Bostick2018}. In comparison with inorganic materials, such organic systems exhibit great flexibility. They can be easily bent, stretched, and twisted \cite{Gore2006, Lionnet2006, Gross2011, Durickovic2013, Lof2019}, allowing a mechanical modulation of the amplitude of electron hopping. It is now clear that in biological molecular systems, electronic transport via long-range hoppings, additional to the nearest-neighbor one, can be significant \cite{Bostick2018}, which already played an important role in generating spin-selective phenomena in the protein and DNA \cite{Guo2014, Mishra2013, Guo2012, Gohler2011}. This motivates to address the influence of long-range electron hoppings on the topological phases of these helical molecules, and the long-range effect was indeed considered in the DNA topological superconductors \cite{Tang2019}. Nevertheless, interesting questions remain as to whether helical organic molecules with long-range hoppings can still implement a Thouless quantum pump and if yes, how its topological nature and current characteristics are modified.

In this paper, we study the topological Thouless pump in helical organic molecules by explicitly take account of the long-range electron hoppings. It is demonstrated that while the adiabatical charge pumping still occurs in the system, the long-range hoppings can induce a topological phase transition with transport features appearing in the pumped current through the molecule. The phase transition is characterized by the closure of a band gap at a critical value of the decay exponent, which separates two topologically distinct phases with strong and weak long-range hoppings. We show that the Chern invariants of the bulk bands adjacent to the gap and the evolution of midgap end states in the pumping process indicate clear differences between these two phases of the helical molecules. In particular, the strong and weak hopping phases support the transport of pumped electrons through the molecule in opposite directions, giving rise to two quantized current plateaus with different signs and amplitudes, as set by the sum over the Chern numbers of all the filled bands below the gap. However, a sharp reversal of the pumped current plateau is absent near the topological phase boundary. It features instead a smooth crossover due to the finite-size effect. Moreover, the experimental observability of these transport characteristics at finite temperatures is also discussed.

The remainder of the paper is organized as follows. Section II introduces the model Hamiltonian for the Thouless pump realized in single-helical organic molecules with long-range hoppings, and explains some necessary details of the computational scheme. Numerical results are presented in Sec.\,III. Sec.\,III A analyses the bulk bands and end states of the system, revealing a topological phase transition induced by the long-range hoppings. We discuss consequences of the topological phase transition in the pumped current in Sec.\,III B. Finally, Sec.\,IV is devoted to a conclusion.

\section{II. Model and Method}
The $\alpha$-helical protein and singe-stranded DNA molecules possess similar helical structures. In the presence of the long-range hoppings and an rotating electric field perpendicular to their helical axes, these single-helical organic molecules are modeled by the following Hamiltonian \cite{Guo2017,Guo2014,Shih2008}
\begin{equation}
{\cal H}(t)= \sum_{n=1}^N \varepsilon_n(t) c_n^{\dagger}c_n+ \sum_{n=1}^{N-1} \sum_{j=1}^{N-n} t_j (c_{n}^{\dagger} c_{n+j} +c_{n+j}^{\dagger} c_{n}).
\end{equation}
Here $c_n^\dagger$ ($c_n$) creates (annihilates) an electron with energy $\varepsilon_n(t)$ at site $n$ and the total number of sites, i.e., the molecular length, is denoted by $N$. The sites represent the amino acids for the protein and the nucleobases for the DNA. $t_j=t_1e^{-\mu(j-1)}$ describes the hopping amplitude between the $n$th site and the $(n+j)$th site \cite{Tang2019,Guo2014}, which exponentially decays with the intersite distance, as characterized by $\mu$ the decay exponent. Note that the previous helical-molecule realization \cite{Guo2017} of topological pumping only took account of the nearest-neighbor hopping $t_1$ (i.e., the case of $\mu\rightarrow\infty$). Since in realistic protein, DNA, and other helical organic molecules the decay exponent $\mu$ is very different, we thus treat $\mu$ as a variable parameter in the model study. This is also inspired by the fact that in experiments $\mu$ can be tuned to a large degree by stretching and twisting the helical molecules and thus changing their intersite distances \cite{Gore2006, Lionnet2006, Gross2011, Durickovic2013, Lof2019}. Effects of the helical structure and the rotating electric field are manifested in the time-dependent on-site energy $\varepsilon_n(t)$, which reads \cite{Guo2012PRB,Pan2015}
\begin{equation}
\varepsilon_n(t)=V_{\rm{g}}\cos\big[n\phi_0 -\varphi(t)\big],
\end{equation}
where $V_{\rm{g}}=eR\mathscr{E}$ describes the half energy difference across the molecules along the direction of the electric field $\mathscr{E}$, $R$ denotes the cross-sectional radius of the molecules, $\phi_0$ is the twist angle between two neighboring sites in the cross section, and the phase $\varphi(t)=2\pi ft$ represents the orientation of the electric field, with $f$ the rotational frequency. In the absence of long-range hoppings, i.e., $t_j=0$ for $j>1$, the Hamiltonian (1) is equivalently a dimension-reduced mapping of the Harper-Hofstadter model \cite{Harper1955,Hofstadter1976} describing a quantum Hall system in two-dimensional square lattice. In this mapping, the twist angle $\phi_0/2\pi$ corresponds to the magnetic flux per primitive cell in units of the flux quantum, and the direction of electric field $\varphi(t)$ is reduced from the transverse momentum. As already shown in Ref.\,\cite{Guo2017}, such helical molecules can be adiabatically pumped by slowly rotating the external electric field and hence modulating the on-site energy periodically in time, which shares the same topological origin as the integer quantum Hall effect \cite{Thouless1982,Kohmoto1985}. The main ingredient of the present work is to investigate consequences of the long-range hopping $t_j$ in this topological Thouless pump.

Our computational scheme is straightforward. Considering $\varphi(t)$ as a time-independent variable parameter $\varphi$, we diagonalize the time-independent Hamiltonian (1) with open boundary conditions to obtain the single-particle eigenenergies $E_l$ and eigenstates $\psi_l$. From this spectrum, the end states and the spatial distribution of electron density $P_{ln}=|\psi_{ln}|^2$ can be identified ($\psi_{ln}$ being the amplitude of the $l$th eigenstate $\psi_l$ at site $n$). Under the periodic boundary condition, the longitudinal momentum $k$ defined within the first Brillouin zone (BZ) is a good quantum number. Diagonalizing the Hamiltonian (1) in the momentum space yields the Bloch state $u_i(k, \varphi)$ corresponding to the $i$th energy band $E_i(k,\varphi)$. In this case, the bulk band topology is characterized by the first Chern number \cite{Thouless1983,Ozawa2019}
\begin{equation}
C_i=-\frac{1}{\pi}\int_\textrm{BZ}\textrm{d}k\int_0^{2\pi}\textrm{d}\varphi\,\,\textrm{Im}\Big\langle\frac{\partial u_i(k,\varphi)}{\partial k}\Big|\frac{\partial u_i(k,\varphi)}{\partial \varphi}\Big\rangle,
\end{equation}
which is a $\mathbb{Z}$ topological invariant.

In order to calculate the current pumped adiabatically by the rotating electric field, the explicit time dependences of the phase $\varphi(t)$ and thus the Hamiltonian (1) need to be considered, resulting in that the eigenenergy $E_l(t)$, the eigenstate $\psi_l(t)$, and the electron density $P_{ln}(t)$ all become time-dependent. We then attach two electrodes to the helical molecule: the left ($L$) lead coupled to the first site and the right ($R$) lead to the last site. The additional Hamiltonian terms are
\begin{equation}
{\cal H}_{0}=\sum_{\mathbf{k},\alpha}\varepsilon_\mathbf{k}c^\dagger_{\mathbf{k}\alpha}c_{\mathbf{k}\alpha}+\sum_\alpha t_0 (c^\dagger_{\mathbf{k}\alpha}c_{n_\alpha}+\textrm{H.c.}),
\end{equation}
where $\alpha=L/R$, $n_L=1$, $n_R=N$, $c^\dagger_{\mathbf{k}\alpha}$ ($c_{\mathbf{k}\alpha}$) creates (annihilates) an electron with momentum $\mathbf{k}$ and energy $\varepsilon_\mathbf{k}$ in the $\alpha$ lead, and $t_0$ is the tunneling amplitude between the leads and the end sites of the molecule. It is convenient to introduce the quantity, $\Gamma=2\pi \rho t_0^2$ with $\rho$ the density of lead states, to represent the lead-molecule coupling strength. Note that there is no bias voltage between the two leads, i.e., they are held at the same chemical potential, having the same Fermi energy $E_F$. The current $I(t)$ flowing from the $L$ lead to the molecule is given by the time derivative of the electron number  $M=\sum_\mathbf{k}c^\dagger_{\mathbf{k}L}c_{\mathbf{k}L}$ of the $L$ lead, $I(t)=-e\frac{\textrm{d}M}{ \textrm{d}t}=-\frac{e}{i\hbar}\langle[M,\,{\cal H}(t)+{\cal H}_0]\rangle$. After some algebraic manipulations \cite{Guo2017, Meir1992}, one obtains $I(t)=\sum_{l=1}^N I_l(t)$ and
\begin{equation}
I_l(t)=\frac{e\Gamma}{\hbar}P_{l1}(t)\Big\{f_{_\textrm{FD}}[E_l(t)]-m_l(t)\Big\},
\end{equation}
where $f_{_\textrm{FD}}(\varepsilon)$ is the Fermi-Dirac distribution function and the electron occupation $m_l(t)$ of the $l$th eigenstate can be determined by self-consistently solving the equation,
\begin{equation}
\frac{\textrm{d}m_l(t)}{\textrm{d}t}=\frac{\Gamma}{\hbar}\big[P_{l1}(t)+P_{lN}(t)\big]\Big\{f_{_\textrm{FD}}[E_l(t)]-m_l(t)\Big\}.
\end{equation}
It is defined that the adiabatically pumped current is the average current over one pump cycle. Therefore, the integral,
\begin{equation}
I_l=f\int_0^{\frac{1}{f}}I_l(t)\,\textrm{d}t,
\end{equation}
gives the pumped current through the $l$th eigenstate and the total pumped current $I=\sum_{l=1}^NI_l$.

\section{III. Results and discussions}
In the numerical results presented below, we fix the nearest-neighbor hopping $t_1=0.1\textrm{eV}$ as the energy unit, $V_{\rm{g}}=1.5$, the twist angle $\phi_0=2\pi/5$, and the molecular length $N=100$, unless stated otherwise. For calculating the pumped current, we take the lead-molecule coupling $\Gamma=10^{-6}$ and the rotational frequency $f=10^5\textrm{Hz}$, such that the adiabatic condition, $hf\ll\Gamma,\,t_1$ ($h$ is the Planck constant), always holds. The temperature $T$ is set to be $k_{\textrm{B}}T=0$ ($k_{\textrm{B}}$ the Boltzmann constant), unless stated otherwise.

\subsection{A. Bulk topology and end states}

\begin{figure}
\includegraphics[width=1.0\columnwidth]{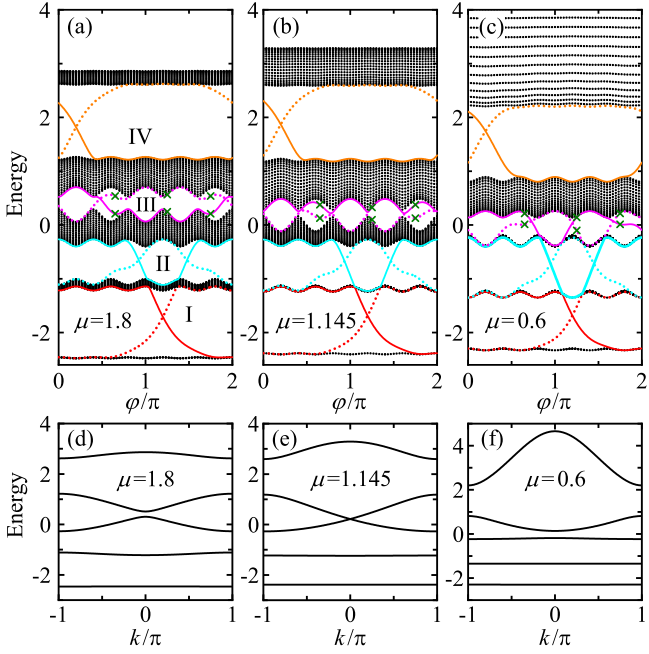}
\caption{\label{fig1} (a)-(c) Energy spectra of single-helical organic molecules with open boundary conditions, for different long-range decay exponents $\mu$, as a function of the phase $\varphi$. The colored lines indicate the evolutions of electronic states traversing energy gaps, and the green crosses mark the electronic states whose spatial distributions are shown in Fig.\,3. (d)-(f) Energy bands of single-helical molecules under periodic boundary conditions for different $\mu$ at fixed $\varphi=0$.}
\end{figure}

Before investigating the consequence of long-range hoppings on the adiabatic pumping, it is helpful to examine its effect on the bulk topology and end states by considering the orientation of electric field $\varphi(t)$ as a time-independent but variable parameter $\varphi$. Figure 1 presents the energy spectra of single-helical organic molecules as functions of the phase $\varphi$ [Figs.\,1(a)-1(c)] and the momentum $k$ [Figs.\,1(d)-1(f)], respectively with open and periodic boundary conditions, for different long-range decay exponents $\mu$. In either case, the molecule possesses five energy bands (bands $1$-$5$, bottom to top) separated by four gaps [gaps I-IV, bottom to top, see, e.g., Fig.\,1(a)], since its unit cell contains five sites due to $\phi_0=2\pi/5$. These energy bands are asymmetric above and below zero energy because the presence of the long-range hoppings breaks the electron-hole symmetry of the system. Interestingly, there is a pair of end states in each energy gap. As the phase $\varphi$ varies, these state pairs, denoted by colored lines in Figs.\,1(a)-1(c), traverse the gaps and intersect at some values of $\varphi$. For odd numbers of intersection points [see gaps I and IV in Figs.\,1(a)-1(c) and gap III in Fig.\,1(c)], the electronic states at $\varphi=0$ and $2\pi$ on the same colored line are different, signaling an evolutionary cycle of $4\pi$ just like a M\"{o}bius strip. When the number of intersection points is even [see gap II in Figs.\,1(a)-1(c) and gap III in Fig.\,1(a)], the evolutionary period is $2\pi$.

The most striking feature lies in the evolution of energy spectra with the variation of the long-range hoppings. It is shown in Fig.\,1 that upon decreasing the decay exponent $\mu$ (i.e., enhancing the long-range hoppings), the system undergoes a topological phase transition at a critical decay exponent $\mu_c\simeq1.145$. For $\mu>\mu_c$ [Figs.\,1(a) and 1(d)], gap III exists and the two midgap end states intersect twice as a function of $\varphi$ with a period of $2\pi$. Gap III closes at $\mu=\mu_c$ [Figs.\,1(b) and 1(e)] and reopens for $\mu<\mu_c$ [Figs.\,1(c) and 1(f)]. In the latter case, the two midgap end states intersect three times as a function of $\varphi$ with a period of $4\pi$. Therefore, the transition at the critical decay exponent separates two topologically distinct phases of the system. Note that gaps I, II, and IV and the $\varphi$-evolutions of the end states in these gaps do not change qualitatively in the whole range of $\mu$.

\begin{figure}
\includegraphics[width=1.0\columnwidth]{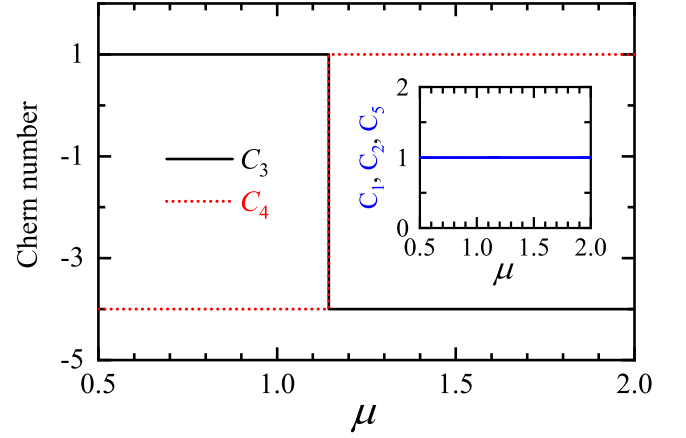}
\caption{Chern numbers, $C_1\sim C_5$, for the five energy bands of the single-helical molecule as a function of the long-range decay exponent $\mu$. }
\end{figure}

This topological phase transition driven by the long-range hoppings is also manifested in the bulk topology characterized by the Chern numbers. As the decay exponent $\mu$ increases past the critical value $\mu_c$, the Chern number $C_3$ of band $3$ jumps from $1$ to $-4$, while the Chern number $C_4$ of band $4$ jumps backwards (Fig.\,2). On the other hand, the Chern numbers of bands $1$, $2$, and $5$ are always $1$ for arbitrary $\mu$ (inset of Fig.\,2), indicating that no phase transition occurs in these energy bands.

\begin{figure}
\includegraphics[width=1.0\columnwidth]{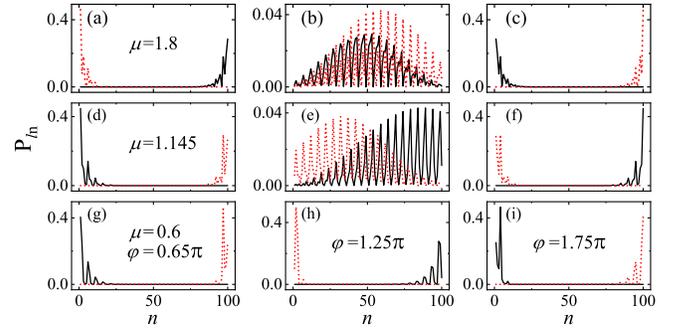}
\caption{Spatial distributions of electron density $P_{ln}$ for $18$ electronic states marked by $18$ green crosses shown in Figs.\,1(a)-1(c), with solid (dotted) lines here corresponding to the solid (dotted) lines there. More specifically, (a)-(c) for $\mu=1.8$ and $\varphi/\pi=0.65,\,1.25,\,1.75$; (d)-(f) for $\mu=\mu_c\simeq1.145$ and $\varphi/\pi=0.65,\,1.25,\,1.75$; (g)-(i) for $\mu=0.6$ and $\varphi/\pi=0.65,\,1.25,\,1.75$.}
\end{figure}

To further elucidate the distinct topological nature of the two phases separated by the transition at $\mu=\mu_c$, we study the spatial distribution of electronic states marked by the two magenta lines pertinent to gap III [see Figs.\,1(a)-1(c)]. Depending on the values of the phase $\varphi$, these states are localized at the left end (L) or right end (R) of the molecule, when the lines are within the energy gap. They can also be extended states (E) distributed in the whole system, when the lines merge into band $3$ or band $4$. More specifically, in the topological phase of weak long-range hoppings ($\mu>\mu_c$), by varying $\varphi$ from $0$ to $2\pi$, the spatial distributions of the states along the solid and dotted magenta lines evolve following
\begin{eqnarray}
\textrm{solid:}& \textrm{E}\rightarrow\textrm{R}\rightarrow\textrm{E}\rightarrow\textrm{L}\rightarrow\textrm{E},\\
\textrm{dotted:}& \textrm{R}\rightarrow\textrm{E}\rightarrow\textrm{L}\rightarrow\textrm{E}\rightarrow\textrm{R},
\end{eqnarray}
with representative distributions shown in Figs.\,3(a)-3(c). In the topological phase of strong long-range hoppings ($\mu<\mu_c$), the evolutions of these states are
\begin{eqnarray}
\textrm{solid:}&\textrm{E}\rightarrow\textrm{L}\rightarrow\textrm{E} \rightarrow\textrm{R}\rightarrow\textrm{E}\rightarrow\textrm{L},\\
\textrm{dotted:}&\textrm{L}\rightarrow\textrm{E}\rightarrow\textrm{R} \rightarrow\textrm{E}\rightarrow\textrm{L}\rightarrow\textrm{E}\rightarrow \textrm{R}\rightarrow\textrm{E},
\end{eqnarray}
and Figs.\,3(g)-3(i) present the spatial distributions at some representative values of $\varphi$. Obviously, the spatial evolutions of the states in gap III are very different between the weak and strong hopping phases. This difference in the spatial distributions of end states as a function of $\varphi$ will have important consequences in the charge pumping effect discussed subsequently. At the critical point $\mu=\mu_c$, although gap III is closed, the electronic states in the magenta lines can still be localized at the molecular ends for some values of $\varphi$ at which the lines do not merge into the bands [Figs.\,3(d)-3(f)].

\subsection{B. Topological charge pumping}

\begin{figure}
\includegraphics[width=1.0\columnwidth]{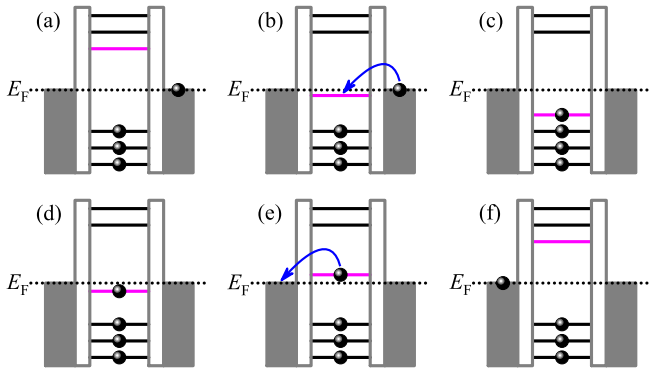}
\caption{Schematic diagrams of the charge pumping process in single-helical organic molecules coupled with two electrodes. (a)-(f) correspond to the phase $\varphi(t)$ evolving in time from $0$ to $2\pi$. The magenta lines here represent the energy levels marked by the solid magenta lines in Fig.\,1(a), and the black balls symbolize electrons.}
\end{figure}

We now turn to investigate the adiabatic charge pumping in the single-helical organic molecules with long-range hoppings. The focus is to demonstrate how the topological phase transition driven by the long-range hoppings is manifested in the pumped current through the system. Taking the end state indicated by the solid magenta line in Fig.\,1(a) as an example in the weak hopping ($\mu>\mu_c$) phase, we illustrate in Fig.\,4 the physical mechanism of electronic transport through topological end states driven by the slowly rotating electric field. Placing the Fermi energy $E_F$ in gap III, we start with the initial state at the zero phase $\varphi(t)=0$. This initial state in the solid magenta line is an extended state which lies above $E_F$ and touches the bottom of band $4$ [Figs.\,1(a) and 4(a)]. Under adiabatic modulation of the pumping parameter $\varphi(t)$ by rotating the electric field, the state evolves with time. Specifically, as $\varphi(t)$ increase from $0$ to $\pi$, the state is dragged downwards from the bottom of band $4$ to the top of band $3$ [Figs.\,4(a)-4(c)] and its spatial distribution evolves according to $\textrm{E} \rightarrow \textrm{R} \rightarrow \textrm{E}$ [Eq.\,(8)]. Within this process, the state is localized at the right end of the molecule while it crosses the Fermi energy. An electron can then tunnel into the molecule from the right electrode [Fig.\,4(b)]. Further adjusting $\varphi(t)$ from $\pi$ to $2\pi$ pushes the state from the top of band $3$ up to the bottom of band $4$ [Figs.\,4(c)-4(f)], and its spatial distribution changes as $\textrm{E}\rightarrow\textrm{L}\rightarrow\textrm{E}$ [Eq.\,(8)]. In particular, the electron is transferred to the left electrode because the state is now localized at the left end of the molecule while passing through the Fermi energy [Fig.\,4(e)]. Therefore, over each pump cycle, an electron is pumped from the right electrode to the left one across the molecule. This is also true for the state marked by the dotted magenta line in gap III. As a result, in the weak hopping ($\mu>\mu_c$) phase, the pumped current carried by the two end states in gap III, as a function of $E_F$, develops a plateau at $-2ef$ [see the magenta line in Fig.\,5(a)], when $E_F$ is in the gap and thus the system behaves as a Chern insulator. The minus ``$-$" here represents the current direction from the left to the right. Remarkably, the index theorem \cite{Chiu2016} relates the number of pumped electrons after one cycle and the sum of Chern invariants of occupied bands below gap III. The current plateau can then be written as $-2ef=\sum_{i=1}^{3}C_ief$, establishing the bulk-boundary correspondence \cite{Ozawa2019,Chiu2016}. Note that $|C_1+C_2+C_3|$ also counts the number of intersections of the two end states in gap III as $\varphi(t)$ runs in $[0,\,2\pi]$.

\begin{figure}
\includegraphics[width=1.0\columnwidth]{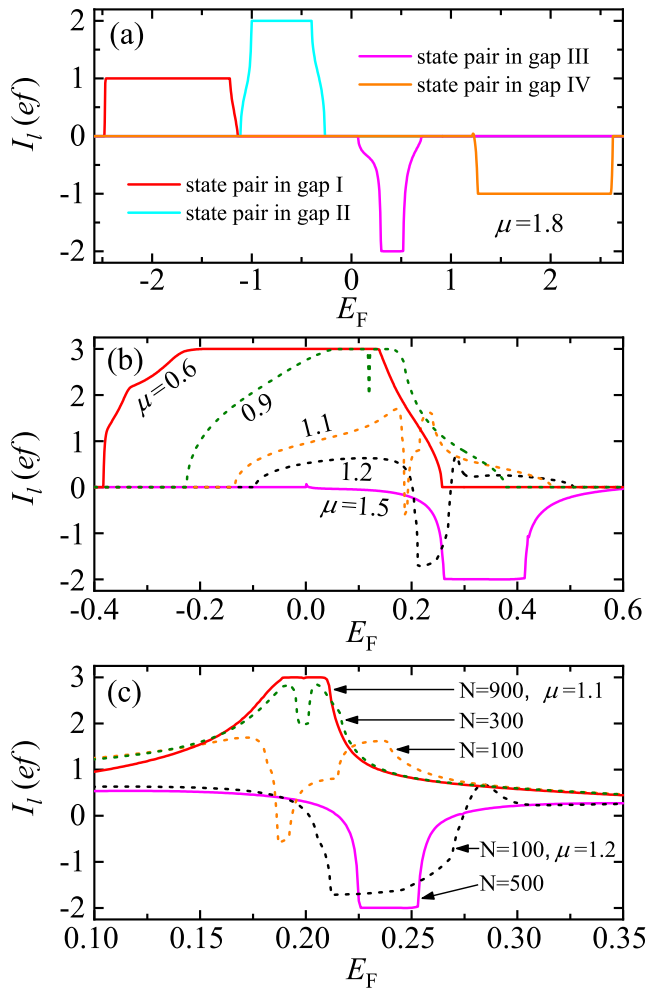}
\caption{(a) Pumped current $I_l$ through end state pairs in gaps I-IV, as a function of the Fermi energy $E_F$ in the leads, at the long-range decay exponent $\mu=1.8$ greater than the critical value $\mu_c\simeq1.145$. (b) Pumped current $I_l$ through the state pair in gap III versus $E_F$, for different $\mu$. (c) Pumped current $I_l$ through the state pair in gap III versus $E_F$ for different molecular lengths $N$, at two decay exponents $\mu$ very close to the critical value $\mu_c$.}
\end{figure}

Similar analyses are equally applicable to the pumped current carried by the topological end states in other bulk gaps. Specifically, in the topological phase of weak hoppings ($\mu>\mu_c$), when the Fermi energy sweeps through gaps I, II, and IV, the sum of Chern numbers of occupied bands is $C_1=1$, $C_1+C_2=2$, and $\sum_{i=1}^4C_i=-1$, respectively. This gives rise to quantized current plateaus of $ef$, $2ef$, and $-ef$, as shown in Fig.\,5(a). The heights of these current plateaus do not change quantitatively even in the strong hopping ($\mu<\mu_c$) phase [see Fig.\,6(c)], because the topological nature of the end state pairs in gaps I, II, and IV is robust against arbitrary amplitudes of the long-range hoppings.

However, the sum of Chern numbers of bands below gap III, $\sum_{i=1}^{3}C_i$, changes abruptly from $-2$ to $3$, as $\mu$ reduces to the $\mu<\mu_c$ regime (Fig.\,3). A current plateau of $3ef$ is thus expected in the strong-hopping topological phase, when the Fermi energy sweeps through gap III [see the red line in Fig.\,5(b)]. Figure 5(b) depicts the pumped current carried by the end state pair in gap III as a function of the Fermi energy $E_F$, for different decay exponents ranging from $\mu>\mu_c$ to $\mu<\mu_c$. It is shown that the current plateau features a continuous crossover from $-2ef$ to $3ef$, rather than a sudden jump typical for phase transitions, due to the finite-size effect \cite{Privman1990,Amit2005}. For decay exponents $\mu$ close to the critical value $\mu_c$ and short molecular lengths $N$, the width of gap III can be comparable to the interval between adjacent energy levels in bulk bands, so that the gap is not well defined and the current plateau is absent. This finite-size effect can be circumvented when energy levels in each bands constitute continuums by increasing the molecular length. Indeed, as demonstrated in Fig.\,5(c), for sufficiently long lengths, the pumped current carried by the end states in gap III always exhibits a well-formed plateau at either $-2ef$ (for $\mu>\mu_c$) or $3ef$ (for $\mu<\mu_c$), even though the decay exponents $\mu$ are very close to the critical value.

\begin{figure}
\includegraphics[width=1.0\columnwidth]{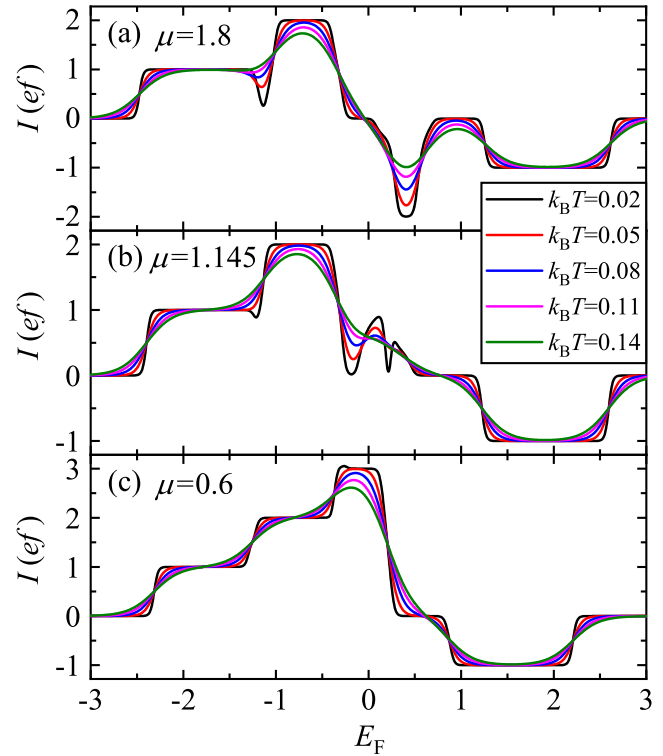}
\caption{Total pumped current $I$ through all end states in the four band gaps as a function of the Fermi energy $E_F$ at different temperatures $T$, for different long-range decay exponents $\mu$ in the $\mu>\mu_c$ phase (a), at the critical point $\mu=\mu_c\simeq1.145$ (b), and in the $\mu<\mu_c$ phase (c), respectively.}
\end{figure}

Obviously, the adiabatic pumping effect is protected by the bulk topology of the molecule. The above transport properties characteristic for the topological phase transition are highly robust against perturbations such as disorder and many-body interaction \cite{Niu1984}. Here, we would like to estimate the observability of these transport characteristics at finite temperatures. Figure 6 presents the temperature evolution of the total pumped current carried by all end states. No matter the system is in the weak hopping ($\mu>\mu_c$) phase [Fig.\,6(a)], in the strong hopping ($\mu<\mu_c$) phase [Fig.\,6(c)], or even at criticality [Fig.\,6(b)], all relevant plateaus are well resolved in the total current when the Fermi energy sweeps consecutively through the bulk gaps. As expected, these plateaus are smeared out with the temperature. Since the plateau widths are roughly given by the corresponding gap widths, the wider the gap, the higher the temperature at which the current plateau can persist. We find that the current plateau at $-2ef$, characteristic for the weak hopping ($\mu>\mu_c$) phase, remains visible at the temperature $k_{\textrm{B}}T=0.02t_1$ [Fig.\,6(a)], while in the strong hopping ($\mu<\mu_c$) phase, the characteristic plateau of $3ef$ can persist up to the temperature $k_{\textrm{B}}T=0.05t_1$ [Fig.\,6(c)]. These give out relatively high temperatures $T\simeq23\,\textrm{K}$ and $58\,\textrm{K}$, if one takes the nearest-neighbor hopping $t_1=0.1\textrm{eV}$ according to the first-principles calculations \cite{Endres2004, Yan2002, Senthilkumar2005, Hawke2010}. Although the exact temperature evolution of the pumped current plateaus depends on the specific model parameters used in this paper, there will be no qualitative changes when other values of the model parameters are used \cite{Guo2017}. Therefore, we expect that the transport signature of the topological phase transition can be observed at finite temperatures, in realistic DNA and protein molecules

\section{IV. Conclusion}
We have studied the topological adiabatic pumping of electrons in single-helical organic molecules with long-range hoppings. It is shown that the strong and weak long-range hopping regimes represent two topologically distinct phases of the system, which is separated by a topological phase transition with characteristic features manifested in the molecular band structure and in the pumped current. In particular, the strong and weak hopping phases support the transport of pumped electrons in opposite directions, resulting in that the quantized plateau in the pumped current as a function of the Fermi energy exhibits different signs and amplitudes between the two phases. Due to the finite lengths of the molecules, we observe a smooth crossover, rather than a sharp reversal, of the current plateau near the topological criticality, which is experimentally accessible in realistic helical molecules. The study of topological physics in helical organic and bioorganic molecular systems is still in its infancy. We hope the present paper could attract more research interest in this direction. Besides being fundamentally interesting on its own right, these studies could also provide novel designing principles for molecular electronic devices by exploiting their topological phases.

\section*{Acknowledgments}
This work is financially supported by NSF-China (Grants No.\,\,11874428, No.\,\,11874187, No.\,\,11921005), the Innovation-Driven Project of Central South University (Grant No.\,\,2018CX044), the National Key R and D Program of China (Grant No.\,\,2017YFA0303301), the Strategic Priority Research Program of Chinese Academy of Sciences (Grant No.\,\,DB28000000), and Beijing Municipal Science \& Technology Commission (Grant No.\,\,Z191100007219013).

\end{document}